\documentclass[aps,prb,tightenlines,floatfix,showpacs]{revtex4}
\usepackage[dvips]{graphicx}
\usepackage[english]{babel}
\usepackage{amsmath}
\usepackage{amssymb}
\usepackage{tensor}

\newcommand{\Ek}{E_{\mathbf{k}}}
\newcommand{\vk}{\mathbf{k}}
\newcommand{\vp}{\mathbf{p}}
\newcommand{\vq}{\mathbf{q}}
\newcommand{\sumk}{\sum_{\mathbf{k}}}

\newcommand{\be}{\begin{eqnarray}}
\newcommand{\ee}{\end{eqnarray}}
\newcommand{\p}{\partial}

\newcommand{\xik}{\xi_{\mathbf{k}}}
\newcommand{\psid}{\psi^{\dagger}}

\allowdisplaybreaks

\begin{document}

\title{Conserving Approximation of Pairing theories in Fermionic superfluid phase}
\author{Yan He$^1$ and Hao Guo$^2$}
\affiliation{$^1$College of Physical Science and Technology,
Sichuan University, Chengdu, Sichuan 610064, China}
\affiliation{$^2$Department of Physics, Southeast University, Nanjing 211189, China}
\date{\today}

\begin{abstract}
Respecting the conservation laws of momentum and energy in a many body theory is very important for understanding the transport phenomena. The previous conserving approximation requires that the self-energy of a single particle can be written as a functional derivative of a full dressed Green's function. This condition can not be satisfied in the $G_0G$ t-matrix or pair fluctuation theory which emphasizes the fermion pairing with a stronger than the Bardeen-Cooper-Schrieffer (BCS) attraction. In the previous work\cite{stressWI}, we have shown that when the temperature is above the superfluid transition temperature $T_c$, the $G_0G$ t-matrix theory can be put into a form that satisfies the stress tensor Ward identity (WI) or local form of conservation laws by introducing a new type of vertex correction. In this paper, we will extend the above conservation approximation to the superfluid phase in the BCS mean field level. To establish the stress tensor WI, we have to include the fluctuation of the order parameter or the contribution from the Goldstone mode. The result will be useful for understanding the transport properties such as the behavior of the viscosity of Fermionic gases in the superfluid phases.
\end{abstract}
\maketitle

\section{Introduction}

For strongly correlated systems, such as ultra-cold Fermi gases in the unitary limit with divergent scattering length, perturbation calculations are not reliable because of the lack of small parameters. To capture the strong fluctuations, various approximating methods have been invented, one important example is the t-matrix theory. The t-matrix theory emphasizes the pairing effect between the fermions due to the stronger-than-BCS attractions. It provides a natural explanation of the preformed pairs before condensation and the formation of the pseudogap. The t-matrix theory has many applications to the ultra-cold Fermi gas because of the experimental tunable interactions.

In recent years a lot of experimental\cite{Thomas} and theoretical works\cite{HaoPaper,Schaefer,Son1,NJOP} in ultra-cold Fermi gases focus on the transport phenomena, such as the behavior of the shear viscosity. In order to give reliable calculations of it, the chosen many body theory should respect the momentum and energy conservation laws. This is the reason that Baym and Kadanoff proposed the conserving approximation long time ago\cite{Baym1,Baym2}. The conserving approximation actually sets up a more stringent consistency requirement to the many body theory. Based on the similar consideration, Haussmann and Zwerger studied a t-matrix theory with two full dressed propagator in the ladder which is also known as $GG$ theory\cite{Haussmann}. There are also two other important types of t-matrix theories. The simplest version of t-matrix theory known as $G_0G_0$ theory with two bare fermion propagator in the ladder was pioneered by Nozieres and Schmitt-Rink\cite{NSR,OurAnnals}. The other one is the $G_0G$ theory first introduced by Kadanoff and Martin\cite{KM}, which is designed to be compatible with the BCS-Leggett generalization\cite{Chen-review,Leggett}. The conserving approximation condition is usually considered not satisfying by these two types of t-matrix theories. However, in our previous work\cite{stressWI} we have shown that certain vertex corrections can be introduced such that the Ward identities of stress tensor or energy-momentum tensor are satisfied both in the $G_0G_0$ and $G_0G$ t-matrix theories above $T_c$, which also means that the momentum and energy are conserved locally. This result opens a way to improve these two t-matrix theories. In this paper, we will show how the local conservation laws are satisfied in the superfluid phase.

Since the way that the conservation law is satisfied in our theory is quite different from that in the traditional conserving approximation, it is worth to make a detailed comparison of both approaches. In general, the local forms of conservation laws are expressed by the operator equations such as $\p_{\mu}j^{\mu}=0$ and $\p_{\mu}T^{\mu\nu}=0$. Inserting these operators into a $n$-particle Green's function, one can find that the divergence of $n$-particle Green's function equals to some contact terms. These are the most original forms of the Ward identities. They must be respected in the exact theory, however they are usually violated in various levels in the approximated theories.

The key point of traditional conserving approximation is that the self-energy of the one-particle Green's function is a functional derivative $\Sigma=\delta\Phi/\delta G$. Here $\Phi$ is a functional of the dressed Green's function $G$, which also appears as an interacting part of the free energy in Luttinger-Ward formalism\cite{LuttingerWard}. In the field theory language, $\Phi$ can also be represented as the 2-particle-irreducible skeleton vacuum diagrams. While in real practical calculations, one has to choose a particular class of diagrams. Due to this truncation, not all the WIs are satisfied in the conserving approximation. What has been proved is that the one-particle Green's functions satisfy the conservation law and the two-particle Green's functions satisfy the thermodynamical consistency. It must be noted that the conservation of the momentum and energy are proved for the whole system rather than a local form. Moreover, certain important symmetries such as the crossing symmetry determined by Pauli principle are also violated. In another approach proposed by de Dominicis and Martin\cite{Dominicis}, the full dressed two-particle scattering vertex has been introduced as a fundamental quantity£¬which is determined by the parquet equations. In this scheme, the crossing symmetry is now respected but the conservation law is not always satisfied. The FLEX approximation proposed by Bickers\cite{Bickers} is roughly equivalent to the combination of the two approaches mentioned above, however there is still some inconsistency such that the vertex derived from the functional derivative of self-energy is not the same as the one used to calculate the self-energy. In summary, the conserving approximation does not mean that all WIs are automatically satisfied. In order to respect crossing symmetry, one has to treat the full vertex on the same footing as the self-energy.

On the other hand, since the $G_0G_0$ or $G_0G$ self-energy can not be written as a functional derivative, it is believed that in these theories the conservation laws are not respected. In our theory, we focus on the satisfaction of stress tensor WI or local conservation instead of pursuing the integral form of conservation law. Moreover, the canonical stress tensor is treated as an external classical field, and we can introduce an external vertex associating with it. The interaction term in the stress tensor serves as a new bare stress tensor vertex. Although the stress tensor WI has a very complicated momentum dependence, we can rearrange the momentum dependence by introducing a new vertex correction corresponding to the self-energy and show that the WI is satisfied by the full vertex. This also implies that the WIs are satisfied for the one and two-particle Green's functions. When we apply the same scheme to the superfluid phase, we have encountered huge difficulties because the effects of Goldstone modes and the pair fluctuation are mixed with each other. In this paper, we will stick to the BCS mean field level as an easy starting point, the work that the pair fluctuation is included will be built on the top of the BCS mean field theory in the future. As we will see later that, even in this level the problem of satisfying stress tensor WI is non-trivial. Although the translational symmetry is not broken, the Goldstone modes of the broken $U(1)$ symmetry plays a very important role in constructing the stress tensor correlation function which satisfies the conservation laws.

\section{Stress tensor Ward Identity (WI) in normal phase}

Since the WI associated with the stress tensor or momentum current is not very familiar to the condensed matter physics community, we first establish these identities for the general scalar fields in this section, and then apply these general formula to the simplest case, the non-interacting Fermi gas. WIs generally represent the continuity equations expressed in terms of Green's functions or correlation functions. They are obtained by sandwiching the operator continuity equations in various time ordered operator products. These WIs must be satisfied in any theory no matter it is exact or approximate. As a beginning, we start with a simple example about the particle number or the $U(1)$ current
conservation which is given by $\frac{\p\rho}{\p t}+\nabla\cdot{\bf J}=0$.
The Ward identity of this example is given by
\be
q_\mu \Gamma^{\mu}(K+Q,K)=G^{-1}(K+Q)-G^{-1}(K),
\label{Jc}
\ee
which is familiar in QED. Here $K\equiv k^{\mu}=(\omega,\vk)$ and $Q\equiv q^{\mu}=(q^0,\vq)$ are the 4-momenta, $G$ is the dressed Green's function and $\Gamma$ is the dressed vertex defined by the equation
\be
&&\langle\psi(x) J^{\mu} (z) \psi^\dagger(y)\rangle\equiv \int G(x,x') \Gamma^{\mu}.
(x',y',z) G (y',y) d^4 x' d^4y',
\ee
where $x\equiv x^{\mu}=(t,\mathbf{x})$ is the 4-coordinate (We work in the unites where $\hbar=c=1$ throughout this paper). More generally, inserting the current conservation continuity equation into a $n$-particle Green's function of some scalar fields, one can find that the divergence of $n$-particle Green's function is not simply zero.
\be
&&\p_{\mu}\langle J^{\mu}(x)\phi(x_1)\cdots\phi(x_n)\rangle
=-\sum_i\delta(x-x_i)\langle\phi(x_1)\cdots\phi(x_n)\rangle.
\ee
The right hand side is often referred as the ``contact terms" which comes from the fact that the time derivative hits on the time order products. We will see these terms also appear in other cases.

The Ward identity associated with the momentum conservation is more complicated but less familiar, and it is why we focus on it in this paper.
According to the Noether's theorem, the canonical stress tensor is given by
\be
T^{\mu\nu}=\sum_a\frac{\p \mathcal{L}}{\p(\p_{\mu}\phi_a)}\p^{\nu}\phi_a-g^{\mu\nu}\mathcal{L},
\label{EM}
\ee
where $\mathcal{L}$ is the Lagrangian density,
$g_{\mu\nu}=(1,-1,-1,-1)$ is the metric tensor and the index $a$ labels different species of fields.

The stress tensor satisfies the conservation law $\p_{\mu}T^{\mu\nu}=0$.
The general WI associated with the correlation function of the stress tensor and other scalar fields in the coordinate space is given by\cite{Coleman}
\be
&&\p_{\mu}\langle T^{\mu\nu}(x)\phi(x_1)\cdots\phi(x_n)\rangle=-\sum_i\delta(x-x_i)\frac{\p}{\p x_i^{\nu}}\langle\phi(x_1)\cdots\phi(x_n)\rangle,
\ee
where the terms on the right-hand-side are also the contact terms as in the current conservation case. The derivative in the contact terms reflect that the momentum is associate with spatial translation.
Applying the above expression to the 3-point correlation function, we get
\be
&&\p_{\mu}\langle T^{\mu\nu}(x)\psid(y)\psi(z)\rangle=-\delta(x-y)\frac{\p}{\p y^{\nu}}\langle\psid(y)\psi(z)\rangle
-\delta(x-z)\frac{\p}{\p z^{\nu}}\langle\psid(y)\psi(z)\rangle.
\ee
Transforming to the momentum space, we find
\be
q_\mu\Gamma^{\mu\nu}(K+Q,K)=k^{\nu}G^{-1}(K+Q)-(k+q)^{\nu}G^{-1}(K)\label{TWI}.
\ee
This is the general form of the WI associated with the stress tensor for both non-interacting and interacting systems, which will be the focus of this paper.
Comparing it to the WI associated with the $U(1)$ current, i.e. Eq.~(\ref{Jc}),
we see that this WI has a subtle momentum dependence, which makes the establishment of it for the
interacting Fermi gases much more difficult than that associated with the $U(1)$ current. To get a warm-up experience of the stress tensor WI, we begin with the non-interacting Fermi gas with the following Lagrangian density
\be
\mathcal{L}=\frac{i}{2}\psid_{\sigma}\p_t\psi_{\sigma}-\frac{i}{2}\p_t\psid_{\sigma}\psi_{\sigma}
-\frac{1}{2m}\p_i\psid_{\sigma}\p_i\psi_{\sigma}+\mu\psid_{\sigma}\psi_{\sigma}.
\ee
Here $\psi_{\sigma}$ is the Fermionic field with $\sigma=\uparrow,\downarrow$ is the spin index, and we have taken the Einstein summation convention and a symmetric form for the time derivative.
The equation of motion is then given by
\be
\frac{\p\mathcal{L}}{\p\psi_{\sigma}}-\p_t\frac{\p \mathcal{L}}{\p(\p_t\psi_{\sigma})}-\p_i\frac{\p \mathcal{L}}{\p(\p_i\psi_{\sigma})}=-i\p_t\psid_{\sigma}+\frac{1}{2m}\p_i^2\psid_{\sigma}+\mu\psid_{\sigma}=0.
\ee
which is just the Schrodinger equation for a non-interacting Fermionic field.

From Eq.~(\ref{EM}), the components of the canonical stress tensor involving
momentum density and momentum current are given by
\be
&&T^{0j}=-(\frac{i}{2}\psid_{\sigma}\p_j\psi_{\sigma}-\frac{i}{2}\p_j\psid_{\sigma}\psi_{\sigma}),\nonumber\\
&&T^{ij}=\frac{1}{2m}(\p_i\psid_{\sigma}\p_j\psi_{\sigma}+\p_j\psid_{\sigma}\p_i\psi_{\sigma})+\delta^{ij}\mathcal{L},
\ee
which satisfy momentum current conservation $\p_tT^{0j}+\p_iT^{ij}=0$. One can see
that in a non-relativistic theory, the momentum density is essentially
the same as the $U(1)$ current
$J^j=T^{0j}/m$.

In general, the stress tensor is not uniquely defined. Different forms
of the stress tensor lead to different forms of WI. The component $T^{ij}$
contains a time derivative which makes the frequency summation quite complicated. We can make use of the
equation of motion to get rid of the time derivative and get a more convenient expression of $T^{ij}$. Hence one
finds
\be
&&T^{0j}=-(\frac{i}{2}\psid_{\sigma}\p_j\psi_{\sigma}-\frac{i}{2}\p_j\psid_{\sigma}\psi_{\sigma}),\nonumber\\
&&T^{ij}=\frac{1}{2m}(\p_i\psid_{\sigma}\p_j\psi_{\sigma}+\p_j\psid_{\sigma}\p_i\psi_{\sigma})-\delta^{ij}\frac{\p_i^2(\psid_{\sigma}\psi_{\sigma})}{4m}.
\ee
The corresponding (bare) vertices are given by
\be
&&\gamma^{0j}(K+Q,K)=k^j+\frac{q^j}{2},\nonumber\\
&&\gamma^{ij}(K+Q,K)=\frac{(k+q)^ik^j+(k+q)^jk^i}{2m}+\delta^{ij}\frac{q^2}{4m}.
\ee
We will stick to this simplified version of vertex in the rest of this paper.
We can also verify the simplified WI associated with these vertices as
\be
&&q_{\mu}\gamma^{\mu j}(K+Q,K)=(k^j+\frac{q^j}2)[G^{-1}_0(K+Q)-G^{-1}_0(K)].
\ee
It is simplified a little bit comparing to the previous version of WI in Eq.(\ref{TWI}).
We are interested in the physical property of the 2-point stress tensor response functions which is useful to the understanding of transport property such as the viscosity. The stress tensor-stress tensor correlation function for any Fermi gas is given by
\be\label{brs}
Q^{\mu j,ab}(x-y)=-i\theta(x^0-y^0)
\langle [T^{\mu j}(x),\,T^{ab}(y)]\rangle.
\ee
By implementing the conservation law of the stress tensor, the divergence of the stress tensor-stress tensor correlation function in the coordinate space is evaluated as
\be
\p_{\mu}Q^{\mu j,ab}(x-y)=-i\delta(x^0-y^0)\langle [T^{0 j}(x),\,T^{ab}(y)]\rangle.
\label{eq:33}
\ee
After performing the Fourier transformation, we find the corresponding expression in the momentum space of the above equation
\be
q_{\mu}Q^{\mu j,ab}(Q)&=&\langle [T^{0 j}(\vq,t),\,T^{ab}(-\vq,t)]\rangle\nonumber\\
&=& \sum_{\vp,\vk}\Big(p+\frac{q}2\Big)^j\gamma^{ab}(K+Q,K)\langle[\psi^{\dagger}_{\sigma\vp}\psi_{\sigma\vp+\vq},\,
\psi^{\dagger}_{\sigma\vk+\vq}\psi_{\sigma\vk}] \rangle \nonumber\\
&=& \sum_{\vk}\Big(k+\frac{q}2\Big)^j\gamma^{ab}(K+Q,K)(n_{\vk}-n_{\vk+\vq}).
\ee
In the specific case of a non-interacting Fermi gas, the divergence of the stress tensor-stress tensor correlation can also be obtained by the diagrammatic method and making use of the vertex WI as
\be
& &q_{\mu}Q_0^{\mu j,ab}(Q)\nonumber\\
&=&\sum_K q_{\mu}\lambda^{\mu j}(K,K+Q)G_0(K+Q)\gamma^{ab}(K+Q,K)G_0(K)\nonumber\\
&=&\sum_K \Big(k+\frac{q}2\Big)^j\Big[G_0(K)- G_0(K+Q)\Big]\gamma^{ab}(K+Q,K),
\ee
which is consistent with the above generic result.

\section{Stress tensor WI in superfluid phase}

Now we move on to the main focus of this paper, we are going to generalize the above results to the BCS superfluid, i.e. to construct the stress tensor-stress tensor correlation function or the full stress tensor vertex which satisfies WI. From the expression of the bare vertex $\gamma^{ij}$, one might guess that a bubble diagram with two vertex insertions may give rise to the correct stress tensor correlation function. However, one can easily verify that this correlation function violates the WI in the superfluid phase. The reason is similar to what happens to the current-current correlation function. In the superfluid pase, the broken $U(1)$ symmetry generates the Goldstone mode, whose contribution cannot be ignored in general gauge choices. Although the translational invariance is not broken, the Goldstone mode still makes important contributions to the stress tensor response theory, as we will see in the following discussions.

For convenience, we use a more compact expression in the Nambu space to discuss our theory. We introduce the two-component Nambu fermion $\Psi^{\dagger}_{\vk}=(\psi_{\uparrow\vk},\psi^{\dagger}_{\downarrow-\vk})$, then the Hamiltonian is
\be
H=\sum_{\vk}\Psi^{\dagger}_{\vk}(\xi_{\vk}\sigma_3-\sigma_1\Delta)\Psi_{\vk},
\ee
where $\xi_{\vk}=\frac{\vk^2}{2m}-\mu$ is the free fermion dispersion, and $\Delta$ is the order parameter given by $\label{Delta}
\Delta(\mathbf{x})=-g\langle\psi_{\uparrow}(\mathbf{x})\psi_{\downarrow}(\mathbf{x})\rangle$ with $g$ being the coupling constant.
It is easy to see that the inverse bare and full matrix Green's functions are
\be
&&\hat{G}^{-1}_0(P)=i\omega_n-\xi_{\mathbf{p}}\sigma_3, \quad \hat{G}^{-1}(P)=i\omega_n-\xi_{\mathbf{p}}\sigma_3+\Delta\sigma_1.
\ee
By direct evaluation, we get the explicit form of the matrix Green's function in the Nambu space
\be
\hat{G}(K)=\left(
             \begin{array}{cc}
               G(K) & F(K) \\
               F(K) & -G(-K)
             \end{array}
           \right).
\ee
Here $G(K)$ and $F(K)$ are the well-known BCS Green's function and anomalous Green's function
\be
&&G(K)=\frac{u^2_{\vk}}{i\omega_n-\Ek}+\frac{v^2_{\vk}}{i\omega_n+\Ek},\\
&&F(K)=-u^2_{\vk}v^2_{\vk}\Big(\frac{1}{i\omega_n-\Ek}-\frac{1}{i\omega_n+\Ek}\Big),
\ee
where the quasi-particle dispersion $\Ek=\sqrt{\xik^2+\Delta^2}$ and coherent factor
$u^2_{\vk},u^2_{\vk}=\frac12(1\pm\xik/\Ek)$. In the Nambu space, the gap equation and number equation can be written as
\be
\Delta=\frac{g}2\sum_K\mbox{Tr}[\sigma_1\hat{G}(K)],\qquad
n=\sum_K\mbox{Tr}[\sigma_3\hat{G}(K)].
\ee
Recall that in the Nambu space the bare external electromagnetic vertex in the Nambu space is given by $\hat{\gamma}^{\mu}(K+Q,K)=(\sigma_3,\frac{\vk+\frac{\vq}{2}}{m})$. Similarly, one can deduce that the bare vertex in the stress tensor linear response theory in the Nambu space becomes
\be
&&\hat{\gamma}^{0j}(K+Q,K)=(k^j+\frac{q^j}{2}),\\
&&\hat{\gamma}^{ij}(K+Q,K)=\Big[\frac{(k+q)^ik^j+(k+q)^jk^i}{2m}+\delta^{ij}\frac{q^2}{4m}\Big]\sigma_3.
\ee
It is straightforward to verify that this bare vertex satisfies the bare WI as follows
\be
q_{\mu}\hat{\gamma}^{\mu j}(K+Q,K)
=(k+\frac q2)^j\Big[\sigma_3\hat{G}^{-1}_0(K+Q)-\hat{G}^{-1}_0(K)\sigma_3\Big].
\ee
To obtain the full stress tensor-stress tensor correlation function which satisfies the WI, we must include the fluctuation of $\Delta$ on the same footing as that of the external metric field which couples to the vertex $\gamma^{\mu j}$. Hence we introduce the generalized external disturbing field
\be
\hat{\mathbf{\Phi}}=\big(\Delta_1,\Delta_2,g_{\nu l}\big)^T
\ee
and the corresponding vertex
\be
\hat{\mathbf{\Lambda}}^1(K+Q,K)
=\Big(\sigma_1,\sigma_2,\hat{\gamma}^{\nu l}(K+Q,K)\Big)^T.
\ee
Then the perturbing Hamiltonian $H'$ can be cast in the scalar product as
\begin{eqnarray}\label{H2}
H'=\sum_{\mathbf{p}\mathbf{q}}\Psi^{\dagger}_{\mathbf{p}+\mathbf{q}}
\hat{\mathbf{\Phi}}\cdot\hat{\mathbf{\Lambda}}^1(K+Q,K)\Psi_{\mathbf{p}}=\sum_{\mathbf{p}\mathbf{q}}\Psi^{\dagger}_{\mathbf{p}+\mathbf{q}}
\hat{\mathbf{\Phi}}_a\hat{\mathbf{\Lambda}}^1_a(K+Q,K)\Psi_{\mathbf{p}}.
\end{eqnarray}
Introducing the imaginary time formalism, the generalized perturbation $\mathbf{\eta}$ of the stress tensor current due to the perturbative Hamiltonian $H'$ is given by
\be
&&\mathbf{\eta}(\tau,\mathbf{q})=\sum_{\mathbf{p}}\langle\Psi^{\dagger}_{\mathbf{p}}(\tau)
\hat{\mathbf{\Lambda}}^2(P+Q,P)\Psi_{\mathbf{p}+\mathbf{q}}(\tau)\rangle,\nonumber\\
&&\mbox{with}\qquad\hat{\mathbf{\Lambda}}^2(K+Q,K)
=\Big(\sigma_1(k+\frac q2)^j,\sigma_2(k+\frac q2)^j,\hat{\gamma}^{\mu j}(K+Q,K)\Big)^T.
\ee
Here $\eta^{\mu}_3=\langle T^{\mu j}\rangle$ denotes the perturbed expectation value of the stress tensor and $\eta_{1,2}$ denotes the perturbed real and imaginary parts of the order parameter multiplied by external momentum. The generalized linear response theory in the momentum space can be written as
\be
\eta_a(Q)=\sum_b Q_{ab}(Q)\Phi_b(Q).
\ee
More explicitly in the matrix form, we have
\be
\left(\begin{array}{c}
\eta_1^j\\
\eta_2^j\\
\eta_3^{\mu j}
\end{array}
\right)=\left(
     \begin{array}{ccc}
       Q_{11}^j & Q_{12}^j & Q_{13}^{j,\nu l} \\
       Q_{21}^j & Q_{22}^j & Q_{23}^{j,\nu l} \\
       Q_{31}^{\mu j} & Q_{32}^{\mu j} & Q_{33}^{\mu j,\nu l}
     \end{array}
   \right)\left(
            \begin{array}{c}
              \Delta_1 \\
              \Delta_2 \\
              g_{\nu l}
            \end{array}
          \right)\label{Qab}
\ee
where the response functions are defined as follows
\begin{eqnarray}
Q_{ab}(\Omega,\vq)=\textrm{Tr}T\sum_{i\omega_n}\sum_{\mathbf{p}} \big(\hat{\Lambda}^2_a(P+Q,P)\hat{G}(P+Q)\hat{\Lambda}^1_b(P,P+Q)\hat{G}(P)\big).
\end{eqnarray}
Detailed expression of these response functions can be found in the appendix.
The component $Q_{33}^{\mu j,\nu l}$ is the generalization of the ``bare'' counterpart given by Eq.~(\ref{brs}). However, it does not satisfy the conservation law since the contribution of the Goldstone mode is not included. Our goal is to find out the ``full'' response kernel which can correctly reflects the spacetime symmetry of the theory, i.e., satisfies the WI.
It can be constructed from the generalized response function $Q_{ab}$ introduced above. Before we do that, we first prove three identities, which are in fact the WIs associated with the response functions since these identities impose the spacetime symmetry on the quantum correlation functions.
\begin{eqnarray}\label{WI}
&&q_{\mu}Q^{\mu j}_{31}=-2i\Delta Q^j_{21},\nonumber\\
&&q_{\mu}Q^{\mu j}_{32}=-2i\Delta Q^j_{22},\nonumber\\
&&q_{\mu}Q^{\mu j,\nu l}_{33}=-2i\Delta Q^{j,\nu l}_{23}+\langle[T^{0j},T^{\nu l}]\rangle.
\end{eqnarray}
To prove them, we need to verify that the matrix BCS Green's function satisfies
\begin{eqnarray}\label{WI0}
(p+\frac q2)^j\Big[\sigma_3\hat{G}^{-1}(P+Q)-\hat{G}^{-1}(P)\sigma_3\Big]
=q_{\mu}\hat{\gamma}^{\mu j}(P+Q,P)+2i\Delta\sigma_2(p+\frac q2)^j.
\end{eqnarray}
This is in fact the WI associated with the Green's function and the stress tensor vertex in the Nambu space. The proof of it is quite straightforward.

For the first identity of Eqs.(\ref{WI}), we have
\begin{eqnarray}\label{WI1}
q_{\mu}Q^{\mu j}_{31}+2i\Delta Q^j_{21}
&=&\textrm{Tr}\sum_P\Big[\big(q_{\mu}\hat{\gamma}^{\mu j}(P+Q,P)
+2i\Delta\sigma_2(p+\frac q2)^j\big)\hat{G}(P+Q)\sigma_1\hat{G}(P)\Big]\nonumber\\
&=&\textrm{Tr}\sum_P\Big[(p+\frac q2)^j\big(\sigma_3\hat{G}^{-1}(P+Q)
-\hat{G}^{-1}(P)\sigma_3\big)\hat{G}(P+Q)\sigma_1\hat{G}(P)\Big]\nonumber\\
&=&\textrm{Tr}\sum_P(p+\frac q2)^j\big[i\sigma_2\hat{G}(P)+\hat{G}(P+Q)i\sigma_2\big]=0
\end{eqnarray}
For the second identity, we have
\begin{eqnarray}\label{WI2}
q_{\mu}Q^{\mu j}_{32}+2i\Delta Q^j_{22}
&=&\textrm{Tr}\sum_P\Big[\big(q_{\mu}\hat{\gamma}^{\mu j}(P+Q,P)
+2i\Delta\sigma_2(p+\frac q2)^j\big)\hat{G}(P+Q)\sigma_2\hat{G}(P)\Big]\nonumber\\
&=&\textrm{Tr}\sum_P\Big[(p+\frac q2)^j\big(\sigma_3\hat{G}^{-1}(P+Q)
-\hat{G}^{-1}(P)\sigma_3\big)\hat{G}(P+Q)\sigma_2\hat{G}(P)\Big]\nonumber\\
&=&-\textrm{Tr}\sum_P(p+\frac q2)^j\big[i\sigma_1\hat{G}(P)+\hat{G}(P+Q)i\sigma_1\big]=0
\end{eqnarray}
For the last one, we have
\begin{eqnarray}\label{WI3}
& &q_{\mu}Q^{\mu j,\nu l}_{33}+2i\Delta Q^{j,\nu l}_{23}\nonumber\\
&=&\textrm{Tr}\sum_P\Big[(p+\frac q2)^j\big(\sigma_3\hat{G}^{-1}(P+Q)
-\hat{G}^{-1}(P)\sigma_3\big)\hat{G}(P+Q)\hat{\gamma}^{\nu l}(P,P+Q)\hat{G}(P)\Big]\nonumber\\
&=&\textrm{Tr}\sum_P(p+\frac q2)^j\Big[\sigma_3\hat{\gamma}^{\nu l}(P+Q,P)\hat{G}(P)
-\hat{G}(P+Q)\hat{\gamma}^{\nu l}(P,P+Q)\sigma_3\Big]\nonumber\\
&=&\sum_{P}(p+\frac q2)^j\textrm{Tr}\big([\hat{G}(P)\sigma_3-\sigma_3\hat{G}(P+Q)]\hat{\gamma}^{\nu l}(P+Q,P)\big)\nonumber\\
&=&\langle[T^{0j}(\vq,t),T^{\nu l}(-\vq,t)]\rangle
\end{eqnarray}

Now we are ready to construct the full stress tensor response function which satisfies the WI. The key point is that the order parameter is not an arbitrary external field but  self-consistently determined. The fluctuation of the phase of the order parameter is the Goldstone mode that plays an important role in the gauge invariant current response theory. It also gives important contribution to the stress tensor response theory. Therefore, the fluctuation of the order parameter must be included in our linear response theory on the same footing as the fluctuation of the external metric field. After imposing the self-consistent condition, the perturbed order parameter can be solved and correctly include the contribution of the Goldstone mode.

The gap equation gives the self-consistent condition $\eta_{1,2}=0$. Applying this relation to Eq.(\ref{Qab}), we find the perturbed order parameter
\begin{eqnarray}\label{D1D2}
\Delta_1=-\frac{Q^{j,\nu l}_{13}Q^j_{22}-Q^{j,\nu l}_{23}Q^j_{12}}{Q^j_{11}Q^j_{22}-Q^j_{12}Q^j_{21}}g_{\nu l},\quad
\Delta_2=-\frac{Q^{j,\nu l}_{23}Q^j_{11}-Q^{j,\nu l}_{13}Q^j_{21}}{Q^j_{11}Q^j_{22}-Q^j_{12}Q^j_{21}}g_{\nu l}.
\end{eqnarray}
After inserting these results into
\begin{eqnarray}\label{CEtemp}
T^{\mu j}=Q^{\mu j}_{31}\Delta_1+Q^{\mu j}_{32}\Delta_2+Q^{\mu j,\nu l}_{33}g_{\nu l},
\end{eqnarray}
we get the usual Kubo expression $T^{\mu i}=K^{\mu i,\nu l}g_{\nu l}$ where the full stress tensor response kernel $K^{\mu j,\nu l}$ is given by
\begin{eqnarray}\label{dK}
K^{\mu j,\nu l}=Q^{\mu j,\nu l}_{33}-\frac{(Q^{j,\nu l}_{13}Q^j_{22}-Q^{j,\nu l}_{23}Q^j_{12})Q^{\mu j}_{31}
+(Q^{j,\nu l}_{23}Q^j_{11}-Q^{j,\nu l}_{13}Q^j_{21})Q^{\mu j}_{32}}
{Q^j_{11}Q^j_{22}-Q^j_{12}Q^j_{21}}.
\end{eqnarray}
where the poles of the second term give the excitations of the collective mode. Hence it only appears in the superfluid phase or more generally the broken-symmetry phase. Now we show that the full stress tensor response kernel does satisfy the conservation law. With the help of the WIs (\ref{WI}), we find that
\begin{eqnarray}\label{CL}
q_{\mu}K^{\mu j,\nu l}&=&-2i\Delta Q^{j,\nu l}_{23}+\langle[T^{0j},T^{\nu l}]\rangle
+2i\Delta\frac{(Q^{j,\nu l}_{13}Q^j_{22}-Q^{j,\nu l}_{23}Q^j_{12})Q^{\mu j}_{21}
+(Q^{j,\nu l}_{23}Q^j_{11}-Q^{j,\nu l}_{13}Q^j_{21})Q^{\mu j}_{22}}
{Q^j_{11}Q^j_{22}-Q^j_{12}Q^j_{21}}\nonumber\\
&=&\langle[T^{0j},T^{\nu l}]\rangle.
\end{eqnarray}
which is also the WI associated with the two-point stress tensor correlation function.

From the expression of $K^{\mu j,\nu l}g_{\nu l}$, we can deduce the full stress vertex as
\be
\hat{\Gamma}^{\mu j}(P+Q,P)=\hat{\gamma}^{\mu j}(P+Q,P)
-(p+\frac q2)^j\sigma_1\Pi^{\mu}_1(Q)-(p+\frac q2)^j\sigma_2\Pi^{\mu}_2(Q)
\ee
where $\Pi_{1,2}$ are defined as
\be
\Pi^{\mu}_1(Q)=\frac{Q^j_{22}Q^{\mu j}_{31}-Q^j_{21}Q^{\mu j}_{32}}{Q^j_{11}Q^j_{22}-Q^j_{12}Q^j_{21}},\quad
\Pi^{\mu}_2(Q)=\frac{Q^j_{11}Q^{\mu j}_{32}-Q^j_{12}Q^{\mu j}_{31}}{Q^j_{11}Q^j_{22}-Q^j_{12}Q^j_{21}}.
\ee
Then the full response kernel can be expressed as
\be
K^{\mu j,\nu l}(Q)=\sum_P\Big[\hat{\Gamma}^{\mu j}(P+Q,P)\hat{G}(P+Q)
\hat{\gamma}^{\nu l}(P,P+Q)\hat{G}(P)\Big]
\ee
By applying the WIs (\ref{WI}), one can see that $\Pi^{\mu}_{1,2}$ satisfies $q_{\mu}\Pi^{\mu}_1(Q)=0$ and $q_{\mu}\Pi^{\mu}_2(Q)=-2i\Delta$. Hence, the full vertex again satisfies the WI
\begin{eqnarray}\label{FGWI}
q_{\mu}\hat{\Gamma}^{\mu}(P+Q,P)=q_{\mu}\hat{\gamma}^{\mu}(P+Q,P)+2i(p+\frac q2)^j\Delta\sigma_2
=(p+\frac q2)^j\Big[\sigma_3\hat{G}^{-1}(P+Q)-\hat{G}^{-1}(P)\sigma_3\Big].
\end{eqnarray}
This is consistent with the conservation law (\ref{CL}) of the response kernel.

\section{conclusion}

In this paper, we construct a stress tensor linear response theory for BCS superfluid which satisfies the WI due to the spacetime symmetry. Therefore we have established the local conservation laws of the energy and momentum in the superfluid phase at the BCS mean-field level. The pairing fluctuation effect is not included for now, while one can treat our theory as a first step toward a the construction of a fully consistent stress tensor linear response theory including the fluctuations from the condensed pairs and the non-condensed
pairs below $T_c$. We emphasize that our conserving approximation is quite different from the $\Phi$-derivable theory in the mathematical form because we are focusing on the local form of the conservation laws or WI. The key point in constructing the full vertex is to treat the phase fluctuation of the order parameter as an independent field which is only constrained by the consistent gap equation. This method is closely parallel to what is used in establishing the $U(1)$ current WI. The result of this paper can be applied to calculate the transport properties such as the viscosity since it can be derived either from the current-current correlations\cite{KM2} or the stress tensor-stress tensor correlations\cite{Luttinger}. To satisfy the WI is crucial to connecting these two different approaches.

Yan He thanks the support by National Natural Science Foundation of China (Grants No. 11404228). Hao Guo thanks the support by National Natural Science Foundation of China (Grants No. 11204032) and Natural Science Foundation of Jiangsu Province, China (SBK201241926).

\appendix

\section{Correlation Functions for BCS Superfluids}\label{app:a}

In this appendix, we list all the detailed expressions of the response functions we have used in the main text.
\be
Q_{11}^{j}(\omega,\vq)&=&\sumk k^j
\Big[(1+\frac{\xik^+\xik^--\Delta^2}{\Ek^+\Ek^-})
\frac{(\Ek^++\Ek^-)[1-f(\Ek^+)-f(\Ek^-)]}{\omega^2-(\Ek^++\Ek^-)^2}\nonumber\\
& &\qquad-(1-\frac{\xik^+\xik^--\Delta^2}{\Ek^+\Ek^-})
\frac{(\Ek^+-\Ek^-)[f(\Ek^+)-f(\Ek^-)]}{\omega^2-(\Ek^+-\Ek^-)^2}\Big]
\ee

\be
Q_{22}^{j}(\omega,\vq)&=&\sumk k^j
\Big[(1+\frac{\xik^+\xik^-+\Delta^2}{\Ek^+\Ek^-})
\frac{(\Ek^++\Ek^-)[1-f(\Ek^+)-f(\Ek^-)]}{\omega^2-(\Ek^++\Ek^-)^2}\nonumber\\
& &\qquad-(1-\frac{\xik^+\xik^-+\Delta^2}{\Ek^+\Ek^-})
\frac{(\Ek^+-\Ek^-)[f(\Ek^+)-f(\Ek^-)]}{\omega^2-(\Ek^+-\Ek^-)^2}\Big]
\ee

\be
Q_{12}^{j}(\omega,\vq)&=&-Q_{21}^{j}(\omega,\vq)=-i\omega\sumk k^j
\Big[(\frac{\xik^+}{\Ek^+}+\frac{\xik^-}{\Ek^-})
\frac{[1-f(\Ek^+)-f(\Ek^-)]}{\omega^2-(\Ek^++\Ek^-)^2}\nonumber\\
& &\qquad-(\frac{\xik^+}{\Ek^+}-\frac{\xik^-}{\Ek^-})
\frac{[f(\Ek^+)-f(\Ek^-)]}{\omega^2-(\Ek^+-\Ek^-)^2}\Big]
\ee

\be
Q_{13}^{j,nl}(\omega,\vq)=\Delta\sumk k^j\gamma^{nl}\frac{\xik^++\xik^-}{\Ek^+\Ek^-}
\Big[\frac{(\Ek^++\Ek^-)[1-f(\Ek^+)-f(\Ek^-)]}{\omega^2-(\Ek^++\Ek^-)^2}
+\frac{(\Ek^+-\Ek^-)[f(\Ek^+)-f(\Ek^-)]}{\omega^2-(\Ek^+-\Ek^-)^2}\Big]
\ee

\be
Q_{31}^{mj}(\omega,\vq)=\Delta\sumk \gamma^{mj}\frac{\xik^++\xik^-}{\Ek^+\Ek^-}
\Big[\frac{(\Ek^++\Ek^-)[1-f(\Ek^+)-f(\Ek^-)]}{\omega^2-(\Ek^++\Ek^-)^2}
+\frac{(\Ek^+-\Ek^-)[f(\Ek^+)-f(\Ek^-)]}{\omega^2-(\Ek^+-\Ek^-)^2}\Big]
\ee

\be
Q_{13}^{j,0l}(\omega,\vq)=\sumk k^jk^l\frac{\Delta\omega}{\Ek^+\Ek^-}
\Big[\frac{(\Ek^+-\Ek^-)[1-f(\Ek^+)-f(\Ek^-)]}{\omega^2-(\Ek^++\Ek^-)^2}
+\frac{(\Ek^++\Ek^-)[f(\Ek^+)-f(\Ek^-)]}{\omega^2-(\Ek^+-\Ek^-)^2}\Big]
\ee

\be
Q_{31}^{0j}(\omega,\vq)=\sumk k^j\frac{\Delta\omega}{\Ek^+\Ek^-}
\Big[\frac{(\Ek^+-\Ek^-)[1-f(\Ek^+)-f(\Ek^-)]}{\omega^2-(\Ek^++\Ek^-)^2}
+\frac{(\Ek^++\Ek^-)[f(\Ek^+)-f(\Ek^-)]}{\omega^2-(\Ek^+-\Ek^-)^2}\Big]
\ee

\be
Q_{23}^{j,nl}(\omega,\vq)=i\sumk k^j\gamma^{nl}\frac{\Delta\omega}{\Ek^+\Ek^-}
\Big[\frac{(\Ek^++\Ek^-)[1-f(\Ek^+)-f(\Ek^-)]}{\omega^2-(\Ek^++\Ek^-)^2}
+\frac{(\Ek^+-\Ek^-)[f(\Ek^+)-f(\Ek^-)]}{\omega^2-(\Ek^+-\Ek^-)^2}\Big]
\ee

\be
Q_{32}^{mj}(\omega,\vq)=-i\sumk \gamma^{mj}\frac{\Delta\omega}{\Ek^+\Ek^-}
\Big[\frac{(\Ek^++\Ek^-)[1-f(\Ek^+)-f(\Ek^-)]}{\omega^2-(\Ek^++\Ek^-)^2}
+\frac{(\Ek^+-\Ek^-)[f(\Ek^+)-f(\Ek^-)]}{\omega^2-(\Ek^+-\Ek^-)^2}\Big]
\ee

\be
Q_{23}^{j,0l}(\omega,\vq)=i\Delta\sumk k^jk^{l}\frac{\xik^+-\xik^-}{\Ek^+\Ek^-}
\Big[\frac{(\Ek^++\Ek^-)[1-f(\Ek^+)-f(\Ek^-)]}{\omega^2-(\Ek^++\Ek^-)^2}
+\frac{(\Ek^+-\Ek^-)[f(\Ek^+)-f(\Ek^-)]}{\omega^2-(\Ek^+-\Ek^-)^2}\Big]
\ee

\be
Q_{32}^{0j}(\omega,\vq)=-i\Delta\sumk k^j\frac{\xik^+-\xik^-}{\Ek^+\Ek^-}
\Big[\frac{(\Ek^++\Ek^-)[1-f(\Ek^+)-f(\Ek^-)]}{\omega^2-(\Ek^++\Ek^-)^2}
+\frac{(\Ek^+-\Ek^-)[f(\Ek^+)-f(\Ek^-)]}{\omega^2-(\Ek^+-\Ek^-)^2}\Big]
\ee

\be
Q_{33}^{mj,nl}(\omega,\vq)&=&\sumk \gamma^{mj}\gamma^{nl}
\Big[(1-\frac{\xik^+\xik^--\Delta^2}{\Ek^+\Ek^-})
\frac{(\Ek^++\Ek^-)[1-f(\Ek^+)-f(\Ek^-)]}{\omega^2-(\Ek^++\Ek^-)^2}\nonumber\\
& &\qquad-(1+\frac{\xik^+\xik^--\Delta^2}{\Ek^+\Ek^-})
\frac{(\Ek^+-\Ek^-)[f(\Ek^+)-f(\Ek^-)]}{\omega^2-(\Ek^+-\Ek^-)^2}\Big]
\ee

\be
Q_{33}^{0j,0l}(\omega,\vq)&=&\sumk k^jk^l
\Big[(1-\frac{\xik^+\xik^-+\Delta^2}{\Ek^+\Ek^-})
\frac{(\Ek^++\Ek^-)[1-f(\Ek^+)-f(\Ek^-)]}{\omega^2-(\Ek^++\Ek^-)^2}\nonumber\\
& &\qquad-(1+\frac{\xik^+\xik^-+\Delta^2}{\Ek^+\Ek^-})
\frac{(\Ek^+-\Ek^-)[f(\Ek^+)-f(\Ek^-)]}{\omega^2-(\Ek^+-\Ek^-)^2}\Big]
\ee

\be
Q_{33}^{0j,nl}(\omega,\vq)=\omega\sumk k^j\gamma^{nl}
\Big[(\frac{\xik^+}{\Ek^+}-\frac{\xik^-}{\Ek^-})
\frac{[1-f(\Ek^+)-f(\Ek^-)]}{\omega^2-(\Ek^++\Ek^-)^2}
-(\frac{\xik^+}{\Ek^+}+\frac{\xik^-}{\Ek^-})
\frac{[f(\Ek^+)-f(\Ek^-)]}{\omega^2-(\Ek^+-\Ek^-)^2}\Big]
\ee

\be
Q_{33}^{mj,0l}(\omega,\vq)=\omega\sumk \gamma^{mj}k^l
\Big[(\frac{\xik^+}{\Ek^+}-\frac{\xik^-}{\Ek^-})
\frac{[1-f(\Ek^+)-f(\Ek^-)]}{\omega^2-(\Ek^++\Ek^-)^2}
-(\frac{\xik^+}{\Ek^+}+\frac{\xik^-}{\Ek^-})
\frac{[f(\Ek^+)-f(\Ek^-)]}{\omega^2-(\Ek^+-\Ek^-)^2}\Big]
\ee

\bibliographystyle{apsrev}
\bibliography{Review2}

\end{document}